\documentclass[a4paper,11pt]{amsart}

\usepackage{amsmath}
\usepackage{amssymb}
\usepackage{amsfonts}
\usepackage{amsthm}
\usepackage{graphicx}

\newcommand{\scalar}[2]{\langle#1\,,#2\rangle}

\newcommand{\norm}[1]{\|#1\|}

\renewcommand{\H}{\mathcal{H}}
\renewcommand{\L}{\mathcal{L}}

\newcommand{\D}{\mathcal{D}}
\renewcommand{\d}{\mathrm{d}}

\newcommand{\C}{\mathcal{C}}

\numberwithin{equation}{section}
\numberwithin{theorem}{section}

\title[Dirac-like operators on the Hilbert space of differential forms]{
Dirac-like operators on the Hilbert space of differential forms on manifolds with boundaries
}

\date{}
\author{Juan Manuel P\'erez-Pardo$^{1,2}$}
\address{$^1$Department of Mathematics,
University Carlos~III, Madrid, Avda. de la Universidad 30, E-28911 Legan\'es
(Madrid), Spain.}
\address{$^2$ INFN-Sezione di Napoli, Via Cintia Edificio 6, I--80126 Napoli, Italy.
}
\email{jmppardo@math.uc3m.es}

\subjclass[2010]{...}
\keywords{Dirac Hamiltonian, Dirac-K\"ahler operator, point-like interactions}

\thanks{The author wants to thank the organisers and scientific committee of ``International Workshop on Quantum Physics: Foundations and Applications 2016" for their invitation and support, where part of this work was developed. The author is supported by a fellowship by the local governement of Region of Madrid, Spain, S2013/ICE-2801. The author was supported also during the development of this work by INFN - Sezione di Napoli. The author is partially supported by grant MTM2014-54692, Ministerio de Economía y Competitividad.
}

\begin{document}

\begin{abstract}
The problem of self-adjoint extensions of Dirac-type operators in manifolds with boundaries is analysed. The boundaries might be regular or non-regular. The latter situation includes point-like interactions, also called delta-like potentials, in manifolds of dimension higher than one. Self-adjoint boundary conditions for the case of dimension 2 are obtained explicitly.
\end{abstract}

\maketitle


\section{Introduction}\label{sec:intro}

In this article we will address several topics related with differential operators defined on the Hilbert space of square integrable functions over differentiable manifolds. The aim is to provide a characterisation of the self-adjoint extensions of the Dirac-K\"ahler operator on manifolds with boundary. While the problem of the Dirac operator with singular potentials has been dealt with in the literature previously, see for instance \cite{Sch72,Sch72b,Wus73,Nen76,EL07}, the study of point interactions for the Dirac operator in dimension higher than one, cf.\ \cite{FG87, BD94} for an analysis in dimension 1, has not been addressed throughly in the literature. This is in contrast with the situation for the Laplace-Beltrami operator where the analysis of point-like interaction has become a field in itself. Recent developments in the effective description of condensed matter systems, cf. \cite{Bal13,Bal16} and references therein, and in particular the developments in graphene physics \cite{SAHR11}, make it relevant to analyse in depth these kind of interactions affecting fermionic fields. The geometric structure inherent to the Dirac-K\"ahler operator eases the characterisation of its domains of self-adjointness. However, the same ideas introduced here could be used to deal with other Dirac type operators.

In order to make this article as self contained as possible we will give first a brief introduction to the notion of Sobolev spaces, while providing the main results and notions that are going to be needed in the latter sections. This will be done in Section \ref{sec:intro}. Section \ref{sec:regular} and Section \ref{sec:nonregular} are devoted to the analysis of regular and non-regular boundaries respectively. The latter providing an example of the characterisation of point-like interactions in dimension 2.

The main objects to study are differential operators acting on functions over differentiable manifolds and the first thing that one has to address in this regard is the fact that differential operators, although linear, are not bounded operators. Therefore,  they cannot be defined over the full Hilbert space. The most paradigmatic example of this situation is the first order derivative, or equivalently, the momentum operator in quantum mechanics, acting on functions on the real line. Indeed, let $\H = \L^2(0,1)$  be the space of square integrable functions over the interval $[0,1]$ and let us denote by $\scalar{\cdot}{\cdot}$ and $\norm{\cdot}=\sqrt{\scalar{\cdot}{\cdot}}$ respectively the scalar product and the associated norm on this space. Now, consider the momentum operator $P = i\frac{\d}{\d x}$\,. Consider the function $\phi(x)= \sqrt{x}$\,. Clearly $\phi\in \L^2(0,1)$.

\begin{equation}
	\scalar{\phi}{\phi} = \int_0^1 x \d x = 1 < \infty\;.
\end{equation}

However,

\begin{equation}
	\scalar{P\phi}{P\phi} = \int_0^1 \frac{1}{4x} \d x \to \infty\;.
\end{equation}
This shows that there are elements of $\H$ such that the momentum operator $P:\H\to\H$ is not defined. The way to solve this difficulty is selecting a domain $\D$ for the operator $P$ which is dense in $\H$, i.e. $\bar{\D}=\H$, and such that $P:\D\to\H$ is well defined for all $\phi\in\D$. For instance, one could pick $\D=\mathcal{C}_c^\infty(0,1)$\,, the space of smooth functions with support at the interior of the interval $(0,1)$ as the domain of the momentum operator. Other choices are possible, for instance $\D=C_c^1(0,1)$\,. Among all the possible choices, the most convenient spaces, cf. \cite{davies:95,reed-simon-1}, for the definition of differential operators are the Sobolev spaces. In the context of a first order differential operator, the Sobolev space to consider is the Sobolev space of order 1. If $\Omega$ is a smooth Riemannian manifold with smooth boundary, the Sobolev space can be defined as

\begin{equation}\label{eq:Sobolev1}
	\H^1(\Omega) = \{\phi\in\L^2(\Omega) | \int_{\Omega} |\frac{\d\phi}{\d x^i}|^2 \d_\mathrm{vol}(\Omega) <\infty,i = 1,\dots,n\}\;,
\end{equation}
where $\{x^i\}$ is any coordinate chart of $\Omega$. In the case that several coordinate patches are needed one has to extend the definition to a covering of the manifold using a partition of the identity. The Sobolev spaces are Hilbert spaces with scalar product given by means of the right hand side of Eq.\ \eqref{eq:Sobolev1}. For further details we refer to \cite{Ad03,Li72}. While this scalar product and the associated norm depend on the choice of the chart, all the norms defined this way are equivalent and thus define the same space of functions. Equivalently, one can define the Sobolev spaces of order $k$ by

\begin{equation}\label{eq:Sobolevk}
	\H^k(\Omega) = \{\phi\in\L^2(\Omega) | \int_{\Omega} |\frac{\partial^{|\vec{k}|}\phi}{\partial^{k_1} x^1\cdots\partial^{k_n} x^n}|^2 \d_\mathrm{vol}(\Omega) <\infty,i = 1,\dots,n\}\;,
\end{equation}
where $\vec{k}=(k^1,\dots,k^n)$ is a multiindex, $|\vec{k}| = \sum_i k^i$ and $\{x^i\}$ is a chart of $\Omega$\,. In these definition the derivatives have to be taken in the weak sense. A particular property of the Sobolev spaces is that the regularity of the functions for a fixed order changes with respect to the dimension of the manifold. For instance, consider the Heaviside step function on $\Omega=[0,1]$ with discontinuity at $x=1/2$\,.

\begin{equation}
	\Theta(x) = 
	    \begin{cases}
	        0 & x<1/2 \\
	        1 & x \geq 1/2
	    \end{cases}\;.
\end{equation}

Clearly $\Theta \in \L^2(0,1)$ but is not an element of $\H^1(0,1)$. A fast argument to show this is that the weak derivative of the Heaviside function is the Dirac-delta with support at $x= 1/2$ and this distribution is not square integrable. In fact, one can prove that $\H^1(0,1)\hookrightarrow\C^0(0,1)$. That is, functions in the Sobolev space of order 1 are continuous functions\footnote{Strictly speaking, since $\H^1(\Omega)$ is a space of equivalence classes of functions, this means that there is at least one continuous representative in each class}.

Let us consider a similar situation with $\Omega = \mathbb{R}^2$, i.e. $\mathrm{dim}\Omega= 2$. Suppose that we have a function that is continuous everywhere except on a region of codimension 1.
A similar argument to that of the previous example shows that functions with this kind of singularities are not allowed in $\H^1(\mathbb{R}^2)$\,. Now, consider the following function given in polar coordinates

\begin{equation}\label{eq:phiinh1}
	\phi(r,\theta) = \log^{1/3}r\;.
\end{equation}

We will just consider integrability on a region around the origin. It is easy to check that

\begin{equation}
	\int_D|\log^{1/3}r|^2r\d r\d\theta < \infty\;,
\end{equation}
where $D$ is any compact region containing the origin in its interior. To prove that this function is in $\H^1(D)$ it is enough to verify the following bounds:

\begin{equation}
	\int_D\left| \frac{\partial \phi}{\partial r} \right|^2 r\d r\d\theta < \infty \quad\text{ and }\quad \int_D\left| \frac{1}{r}\frac{\partial \phi}{\partial \theta} \right|^2 r\d r\d\theta < \infty\,.
\end{equation}
Since $\frac{\partial \phi}{\partial \theta} = 0$ and $\frac{\partial\phi}{\partial r} = \frac{1}{3} \log^{-2/3}r$ the above inequalities hold. And therefore, the function of Eq.\ \eqref{eq:phiinh1} is in $\H^1(D)$. This shows that functions on $\mathbb{R}^2$ that are not continuous at just one point are also allowed to be in the Sobolev space of order 1. In fact, if one wants to recover continuity one has to raise the order of the Sobolev space and make it equal to the dimension of the manifold. These kind of results and many other relations between Sobolev spaces are known as Soboloev embeddings.

Let us do a final remark about the singularities allowed in the Sobolev space of order 1. The example above was provided by a function that was rotationally invariant around the singular point. Let us consider a function of the form $\phi(r,\theta) = f(r)e^{in\theta}$, $n\in\mathbb{Z}$ and $f(r)$ some measurable function. In order for this function to be in $\L^2(\mathbb{R}^2)$ one needs that $f\in\L^2((0,\infty);r\d r)$\,. However, if one wants it to be in the Sobolev space of order 1, the convergence of the integral

\begin{equation}
	\int_D\left| \frac{1}{r}\frac{\partial \phi}{\partial \theta} \right|^2 r\d r\d\theta
\end{equation}
imposes stronger conditions on the regularity at the origin. Namely

\begin{equation}
	\int_0^\infty\frac{1}{r}\left| f(r) \right|^2\d r < \infty \;,
\end{equation}
which implies that $\phi(r,\theta)$ has to be continuous at zero with value $0$ for all $n\neq0$. Hence, if the function is not continuous at a point, but is an element of $\H^1(\mathbb{R}^2)$, there cannot be any angular dependence in a neighbourhood around the singularity.


\section{Self-adjoint extensions of the Dirac-K\"ahler operator on manifolds with regular boundary}\label{sec:regular}

In this section we will consider the Dirac-K\"ahler operator over a Riemannian manifold with regular boundary. We will say that a boundary is regular if it is a smooth submanifold of codimension 1. Let us introduce briefly the Dirac-K\"ahler operator. This will also serve to fix the notation. Let $(\Omega,g)$ be a Riemannian manifold with metric g and let $\Lambda^*(\Omega)$ denote the space of exterior differential forms on the manifold. In general, the elements of $\Lambda^*(\Omega)$ are not of fixed degree and we will denote them as

\begin{equation}\label{eq:genericforms}
\Lambda^*(\Omega) \ni \alpha = \alpha^{(0)}+\alpha^{(1)}+\alpha^{(2)}+\cdots+\alpha^{(n)}\,,
\end{equation}
where $n=\mathrm{dim}(\Omega)$ and $\mathrm{deg}(\alpha^{(k)})=k$\,, i.e. $\alpha^{(k)}$ is a form of degree $k$ or $k$-form. A form of degree $k$ can be expressed in terms of a local chart $\{x^i\}$ as

\begin{equation}
	\Lambda^k(\Omega) \ni \alpha^{(k)} = \sum_{i_1<\cdots<i_k} \alpha^{(k)}_{i_1\cdots i_k}(\mathbf{x})\d x^{i_1}\wedge\dots\wedge\d x^{i_k}\;.
\end{equation}
Eventually, in the special cases of low dimension the components of the forms may be labelled by the coordinates themselves. For instance, 

$$\Lambda^1(\mathbb{R}^3) \ni \alpha^{(1)} = \alpha^{(1)}_x\d x + \alpha^{(1)}_y\d y + \alpha^{(1)}_z\d z\;.$$
The space of exterior forms, which is the space of sections of the vector bundle $\Lambda^*(\Omega)$\,, has a natural Hermitean scalar product. Let us define it first for forms of the same degree $\alpha^{(k)}, \beta^{(k)} \in \Lambda^{k}(\Omega)$:

\begin{equation}
	\scalar{\alpha^{(k)}}{\beta^{(k)}}_k = \int_{\Omega} \bar{\alpha}^{(k)}\wedge\star\beta^{(k)}\;,
\end{equation}
where $\star$ is the Hodge star operator defined on $\Omega$ and $\bar{\alpha}^{(k)}$ is the complex conjugate of the form $\alpha^{(k)}$\,. This scalar product among $k$-forms can be extended to the vector bundle $\Lambda^*(\Omega)$ by considering that forms of different degree are orthogonal. Thus, for $\alpha, \beta \in \Lambda^*(\Omega)$ we have that

\begin{equation}
	\scalar{\alpha}{\beta} =\sum_{k=0}^n\scalar{\alpha^{(k)}}{\beta^{(k)}}_k\;.
\end{equation}
Now, one can define the space of square integrable sections on $\Lambda^*(\Omega)$ in the following way:

\begin{equation}\label{eq:squareintegrablesections}
	\L^2(\Lambda^*(\Omega)) = \{\alpha\in\Gamma(\Lambda^*(\Omega)) \mid \alpha^{(0)}(\mathbf{x}),\dots, \alpha^{(k)}_{i_1\cdots i_k}(\mathbf{x}), \dots,\alpha^{(n)}_{1\cdots n}(\mathbf{x}) \in \L^2(\Omega)\}\;.
\end{equation}
This space, together with the scalar product defines a Hilbert space. 

The most natural differential operator acting on the space of exterior forms is the exterior derivative $\d$, \cite{Ma01}. We shall consider that $\d$ is an operator that can be split according to the degrees of the forms. Thus, $\d = \sum_{k=0}^n\d^{(k)}$, with $\d^{(k)}:\Lambda^{(k)}\to\Lambda^{(k+1)}$, $\d^{(n)}=0$ and $\d^{(k)}$ is such that it is the zero operator when acting on forms of degree different from $k$. A natural question to ask is what is the adjoint of this operator on $\L^2(\Lambda^k(\Omega))$\,. Let us define a differential operator as follows: 

$$\delta^{(k+1)}:\Lambda^{k+1}(\Omega)\to\Lambda^{k}(\Omega)$$
$$\delta^{(k+1)}:= (-1)^{nk+1}\star\d^{(n-k-1)}\star$$
and $\delta^{(k)}=0$ when acting on forms of degree different from $k$. It is easy to check, using the properties of the exterior differential, that

\begin{equation}\label{eq:greenk}
	\scalar{\alpha^{(k)}}{\d^{(k-1)}\beta^{(k-1)}}_k = \scalar{\delta^{(k)}\alpha^(k)}{\beta^{(k-1)}}_{k-1} + \int_\Omega\d^{(n-1)}(\beta^{(k-1)}\wedge\star\bar{\alpha}^{(k)})\;.
\end{equation}
One can now define the operator $\delta=\sum_{k=0}^n\delta^{(k)}$\,. Notice that the second term on the right hand side of Eq.\ \eqref{eq:greenk} is a boundary term. Indeed, if the boundary is regular, one can use Stokes' Theorem to get  an integral along the boundary manifold. Thus, if $\Omega$ has no boundary then $\delta$ is the \emph{formal} adjoint of $\d$\,. It is called the codifferential operator.

As it was shown in Section \ref{sec:intro}, in order to have these operators well defined as operators on the Hilbert space of square integrable sections of $\Lambda^*(\Omega)$\,, one needs to choose an appropriate domain. In the same way that one introduces the space of square integrable sections, one can consider the Sobolev space of sections of order 1 on the vector bundle $\Lambda^*(\Omega)$\,. The latter is defined as 

\begin{equation}
	\H^1(\Lambda^*(\Omega)) =  \{\alpha\in\Gamma(\Lambda^*(\Omega)) \mid \alpha^{(0)},\dots, \alpha^{(k)}_{i_1\cdots i_k}(\mathbf{x}), \dots,\alpha^{(n)}_{1\cdots n}(\mathbf{x}) \in \H^1(\Omega)\}\;.
\end{equation}

It can be proved that, on manifolds without boundary, the adjoint operator of $\d:\H^1(\Lambda^*(\Omega))\to \L^2(\Lambda^*(\Omega))$ is precisely

\begin{equation}
	\delta:\H^1(\Lambda^*(\Omega))\to \L^2(\Lambda^*(\Omega))\;.
\end{equation}
The Dirac-K\"ahler operator can be defined as the sum of these two operators

\begin{equation}\label{eq:DK}
	D=\d + \delta : \H^1(\Lambda^*(\Omega))\to \L^2(\Lambda^*(\Omega))\;,
\end{equation}
which on manifolds without boundary defines a self-adjoint operator in the space of square integrable sections. 

In general, in a manifold with boundary, one has the identity

\begin{equation}\label{eq:GreenDirac}
	\scalar{\alpha}{D\beta}-\scalar{D\alpha}{\beta} = \int_\Omega \d(\beta\wedge\star\bar{\alpha}) - \int_\Omega \d(\bar{\alpha}\wedge\star\beta)\;.
\end{equation}
Despite of what it looks like, the right hand side is not identically zero. This is due to the fact that $\alpha^{(k)}\wedge\star\beta^{(j)}\neq\beta^{(j)}\wedge\star\alpha^{(k)}$ when $k\neq j$\,. This means that the Dirac-K\"ahler operator with domain $\H^1(\Lambda^*(\Omega))$ is not a self-adjoint operator in general. The most straightforward way to define self-adjoint domains is identifying maximally isotropic subspaces of the boundary terms, cf. \cite{Ko75,AIM05,BGP08}\,. We will follow the ideas in \cite{AIM05} to show how to achieve this, for this operator, by means of a simple example. Consider that $\Omega = [0,\infty)$\,. A generic differential form can be expressed by

\begin{equation}
	\alpha = \alpha^{(0)}(x) + \alpha^{(1)}_x(x)\d x
\end{equation}
 and its hodge dual takes then the form

\begin{equation}
	\alpha =  \alpha^{(1)}_x(x) + \alpha^{(0)}(x)\d x \;.
\end{equation}
The right hand side of Eq.\ \eqref{eq:GreenDirac} becoms in this case:

$$\beta^{(0)}(0)\bar{\alpha}^{(1)}_x(0) - \bar{\alpha}^{(0)}(0)\beta^{(1)}_x(0)\;.$$
Maximally isotropic subspaces of this boundary term  can be expressed in an implicit way using the Cayley transform, cf. \cite{AIM05, BGP08, ibortlledo14b}. In this particular case the implicit equation becomes

\begin{equation}\label{eq:asorey}
	\alpha^{(0)}(0)-i\alpha_x^{(1)}(0) = U[	\alpha^{(0)}(0)+i\alpha_x^{(1)}(0)]\;,
\end{equation}
where $U\in\mathcal{U}(1)$. Equivalently, if we parametrise $U=e^{i\lambda}$ for $\lambda\in[0,2\pi)$\,, the above equation can be written as

$$\alpha_x^{(1)}(0) = \tan\frac{\lambda}{2} \alpha^{(0)}(0)\;,\quad\lambda\neq \pi\;,$$
or
$$\alpha^{(0)}(0) = \cot\frac{\lambda}{2}\alpha_x^{(1)}(0) \;,\quad\lambda\neq 0\;.$$

The limiting cases $\lambda=0$ and $\lambda=\pi$ coincide with the situations

$$\lambda = \pi \Rightarrow 
	\begin{cases}
		\alpha^{(0)}(0)= 0 \\
		\alpha^{(1)}(0)\quad \text{Free / no condition}
	\end{cases}\;,
$$
$$\lambda = 0 \Rightarrow 
	\begin{cases}
		\alpha^{(0)}(0)\quad \text{Free / no condition} \\
		\alpha^{(1)}(0)= 0
	\end{cases}\;.
$$
The implicit equations to describe the boundary conditions in the general case can be constructed applying the same ideas. There is only one further comment to make. In manifolds with dimension higher than one, the spaces of boundary data for the Dirac-K\"ahler operator are Sobolev spaces of fractional order, cf. \cite{Ad03}, namely $\H^{1/2}(\partial\Omega)$\,. As occurs for the Laplace-Beltrami operator, cf. \cite{ibortlledo14b}, this imposes certain extra regularity conditions on the unitary operators defining the implicit equation \eqref{eq:asorey}.


\section{Self-adjoint extensions of the Dirac-K\"ahler operator on manifolds with non-regular boundary}\label{sec:nonregular}

In the former section we used Stokes' Theorem to find parameterisations of different domains where the Dirac-K\"ahler operator is self-adjoint. The aim of this section is to provide some tools in order to address the problem when the boundaries are such that Stokes' Theorem does not apply. We will call such boundaries non-regular. It is not intended here to give a full theory. These would require a deep analysis on the different non-regular boundaries that may appear in smooth manifolds and goes beyond the scope of the present article. Instead, we aim to show how can one deal with these non-regular boundaries by means of a particular example. In dimension 1, all the boundaries of smooth manifolds are regular, therefore the lowest dimension where this problem appears is dimension 2. As a guiding example we will consider $\Omega = \mathbb{R}^2\backslash\{0\}$\,. The boundary is in this case $\partial\Omega = \{0\}$\,. The different self-adjoint domains that will be obtained in this situation will be associated with the possible point-like interactions admissible in this system. It is going to be convenient to consider  coordinates adapted to this singularity and therefore we will use polar coordinates centred at $0\in\mathbb{R}^2$\,. In these coordinates the differential forms are parameterised as
\begin{equation}
	\alpha = \alpha^{(0)}(r,\theta) + \alpha^{(1)}_r(r,\theta)\d r + \alpha^{(1)}_\theta(r,\theta)\d \theta + \alpha^{(2)}(r,\theta)\d r\wedge\d \theta\;.
\end{equation}
Expression \eqref{eq:GreenDirac} holds also for this case. In polar coordinates, the right hand side becomes

\begin{equation}\label{eq:boundarytermssingular}
	\int_\Omega \d(\beta^{(0)}\wedge\star\bar{\alpha}^{(1)}) + \int_\Omega \d(\beta^{(1)}\wedge\star\bar{\alpha}^{(2)}) - \int_\Omega \d(\bar{\alpha}^{(0)}\wedge\star\beta^{(1)}) - \int_\Omega \d(\bar{\alpha}^{(1)}\wedge\star\beta^{(2)})\;.
\end{equation}
All these integrals have a similar form, so lets us concentrate on the first one. Since Stokes' Theorem cannot be applied to compute the integral, one needs to compute it by different means. Let us regularise around the singularity. The first term of Eq.\ \eqref{eq:boundarytermssingular} can be split as follows:

\begin{equation}\label{eq:boundarytermsplit}
	\int_\Omega \d(\beta^{(0)}\wedge\star\bar{\alpha}^{(1)}) = \int_{D_\epsilon} \d(\beta^{(0)}\wedge\star\bar{\alpha}^{(1)}) + \int_{\Omega-D_\epsilon} \d(\beta^{(0)}\wedge\star\bar{\alpha}^{(1)})\;,
\end{equation}
where $D_\epsilon$ is a disk of radious $\epsilon$ centred at the origin and with the origin removed. The left hand side  does not depend on $\epsilon$\,. Hence, if the limit for $\epsilon\to0$ exists for both integrals at the right hand side separately, the sum of these limits will be equal to the integral on the left hand side. Notice that because of Eq.\ \eqref{eq:greenk}

\begin{align*}
	\left|\int_{D_\epsilon} \d(\beta^{(0)}\wedge\star\bar{\alpha}^{(1)})\right| &= |\scalar{\alpha^{(1)}}{\d \beta^{(0)}}_{D_\epsilon}-\scalar{{\delta\alpha^{(1)}}}{\beta^{(0)}}_{D_\epsilon}|\\
	    &\leq \mathrm{vol}(D_\epsilon)\left(\norm{\alpha^{(1)}}\norm{\d \beta^{(0)}} + \norm{\delta\alpha^{(1)}}\norm{\beta^{(0)}}\right)\;,
\end{align*}
where the last inequality follows by Cauchy-Schwarz, the fact that $\alpha$ and $\beta$ are in $\H^1(\Lambda^*(\Omega))$ and that $D_\epsilon\subset\Omega$\;. Hence, in the limit $\epsilon\to0$ this integral vanishes. The second term at the right hand side of Eq.\ \eqref{eq:boundarytermsplit} is an integral over a domain with a regular boundary and thus one can use Stokes' theorem to get

\begin{align*}
    \int_{\Omega-D_\epsilon} \d(\beta^{(0)}\wedge\star\bar{\alpha}^{(1)}) 
            &= \int_{\partial D_{\epsilon}} \mathrm{i}^*\left( \beta^{(0)}\bar{\alpha}^{(1)}_r  r \d\theta - \beta^{(0)}\bar{\alpha}^{(1)}_\theta\frac{1}{r}\d r \right)\\
            &=\int_{\partial D_{\epsilon}} \mathrm{i}^*\left(\beta^{(0)}\bar{\alpha}^{(1)}_r  r \d\theta\right)\\
            &=\int_{\partial D_{\epsilon}} \epsilon \beta^{(0)}(\epsilon,\theta)\bar{\alpha}^{(1)}_r(\epsilon,\theta)  r \d\theta\;,
\end{align*}
where $\mathrm{i}:\partial D_\epsilon \to \Omega-D_\epsilon$ is the natural embedding of the boundary and the orientation of $\partial D_\epsilon$ has been chosen such that the integral is performed in the counter-clockwise direction. The second term at the first line vanishes identically. Now, we have to compute the latter integral. In order to do so, we are going to use the Fourier expansions in the variable $\theta$\,.

\begin{equation}
	\beta^{(0)}(\epsilon,\theta) = \sum_{k\in\mathbb{Z}} \beta^{(0)}_k(\epsilon)e^{ik\theta}\;,\quad \alpha^{(1)}_r(\epsilon,\theta) = \sum_{k\in\mathbb{Z}} \alpha^{(1)}_{r,\,k}(\epsilon)e^{ik\theta}\;.
\end{equation}
Substituting in the previous expression we get

\begin{equation}\label{eq:integralfourier}
    \int_{\Omega-D_\epsilon} \d(\beta^{(0)}\wedge\star\bar{\alpha}^{(1)}) = \epsilon\beta^{(0)}_0(\epsilon)\bar{\alpha}^{(1)}_{0,\,r}(\epsilon) + \epsilon \sum_{k\neq0}\beta^{(0)}_k(\epsilon)\bar{\alpha}^{(1)}_{k,\,r}(\epsilon)\;.
\end{equation}
Since $\beta^{(0)}\,,{\alpha}^{(1)}_r \in \H^1(\Omega)$ and from the discussion in Section 1 on the regularity of the functions in the Sobolev space of order 1, the second term in Eq.\ \eqref{eq:integralfourier} has to vanish since there cannot be angular dependence near the singularity at $r=0$\,. Hence, we have finally:

\begin{equation}
	\int_{\Omega} \d(\beta^{(0)}\wedge\star\bar{\alpha}^{(1)}) = \lim_{\epsilon\to0}\epsilon\beta_0^{(0)}(\epsilon)\bar{\alpha}^{(1)}_{r,\,0}(\epsilon)\;.
\end{equation}
The other integrals in Eq.\ \eqref{eq:boundarytermssingular} can be reduced to similar expressions. For instance,
$$\lim_{\epsilon\to0}\frac{1}{\epsilon}\beta_{\theta,\, 0}^{(1)}(\epsilon)\bar{\alpha}^{(2)}_0(\epsilon)\;.$$
The resulting boundary term has the form:
\begin{equation}
\lim_{\epsilon\to0}\left[ \epsilon\beta_0^{(0)}(\epsilon)\bar{\alpha}^{(1)}_{r,\,0}(\epsilon) +  \frac{1}{\epsilon}\beta_{\theta,\, 0}^{(1)}(\epsilon)\bar{\alpha}^{(2)}_0(\epsilon) -\epsilon\bar{\alpha}_0^{(0)}(\epsilon)\beta^{(1)}_{r,\,0}(\epsilon) -  \frac{1}{\epsilon}\bar{\alpha}_{\theta,\, 0}^{(1)}(\epsilon)\beta^{(2)}_0(\epsilon)\right]\;.
\end{equation}
Hence, the self-adjoint extensions of the Dirac-K\"ahler operator in this situation can be obtained by finding maximally isotropic subspaces of the boundary form defined by the limiting values at the singularity. As done in Section \ref{sec:regular}, cf. \cite{AIM05,BGP08,ibortlledo14b}, one can do a change of variables to diagonalise the quadratic form. In this case the transformation takes the form:

\begin{equation}
	\begin{pmatrix}
		\beta^{(0)}_0(\epsilon)\\
		\epsilon \beta^{(1)}_{r,\,0}(\epsilon)\\
		\beta^{(1)}_{\theta,\,0}(\epsilon)\\
		\frac{1}{\epsilon}\beta_0^{(2)}(\epsilon)
	\end{pmatrix}
	\to
	\begin{pmatrix}
		\beta^{(0)}_0(\epsilon) -i \epsilon \beta^{(1)}_{r,\,0}(\epsilon)  \\
		\beta^{(1)}_{\theta,\,0}(\epsilon) - \frac{i}{\epsilon} \beta_0^{(2)}(\epsilon)\\
		\beta^{(0)}_0(\epsilon) + i \epsilon \beta^{(1)}_{r,\,0}(\epsilon)\\
		\beta^{(1)}_{\theta,\,0}(\epsilon) +\frac{i}{\epsilon}\beta_0^{(2)}(\epsilon)
	\end{pmatrix}
\end{equation}
from which it follows that the self-adjoint domains can be parameterised by the group $\mathcal{U}(2)$ in an implicit way by:

\begin{equation}
	\lim_{\epsilon\to0}
	\begin{pmatrix}
		\beta^{(0)}_0(\epsilon) -i \epsilon \beta^{(1)}_{r,\,0}(\epsilon)  \\
		\beta^{(1)}_{\theta,\,0}(\epsilon) - \frac{i}{\epsilon} \beta_0^{(2)}(\epsilon)
	\end{pmatrix} = U
	\lim_{\epsilon\to0}
	\begin{pmatrix}
		\beta^{(0)}_0(\epsilon) + i \epsilon \beta^{(1)}_{r,\,0}(\epsilon)\\
		\beta^{(1)}_{\theta,\,0}(\epsilon) +\frac{i}{\epsilon}\beta_0^{(2)}(\epsilon)
	\end{pmatrix}\;,
\end{equation}
with $U\in\mathcal{U}(2)$\;.


\end{document}